\documentclass[aps,prb,twocolumn,superscriptaddress,showpacs]{revtex4}
\usepackage{graphicx}
\usepackage{graphicx,epstopdf,color}
\usepackage{amsfonts}
\usepackage{amsmath,amssymb,mathrsfs}
\usepackage{bm}
\usepackage{float}
\usepackage{pst-grad}
\usepackage{dcolumn}

\begin{document}

\title{Measurement of the elastoresistivity coefficients of the underdoped iron-arsenide Ba(Fe$_{0.975}$Co$_{0.025}$)$_2$As$_2$}

\author{Hsueh-Hui Kuo}
\affiliation{Stanford Institute for Materials and Energy Sciences, SLAC National Accelerator Laboratory,\\ 2575 Sand Hill Road, Menlo Park, CA 94025, USA} 
\affiliation{Geballe Laboratory for Advanced Materials and Department of Materials Science and Engineering, Stanford University, USA}
\author{Maxwell C. Shapiro}
\affiliation{Stanford Institute for Materials and Energy Sciences, SLAC National Accelerator Laboratory,\\ 2575 Sand Hill Road, Menlo Park, CA 94025, USA} 
\affiliation{Geballe Laboratory for Advanced Materials and Department of Applied
Physics, Stanford University, USA}
\author{Scott C. Riggs}
\affiliation{Stanford Institute for Materials and Energy Sciences, SLAC National Accelerator Laboratory,\\ 2575 Sand Hill Road, Menlo Park, CA 94025, USA} 
\affiliation{Geballe Laboratory for Advanced Materials and Department of Applied
Physics, Stanford University, USA}
\author{Ian R. Fisher}
\affiliation{Stanford Institute for Materials and Energy Sciences, SLAC National Accelerator Laboratory,\\ 2575 Sand Hill Road, Menlo Park, CA 94025, USA} 
\affiliation{Geballe Laboratory for Advanced Materials and Department of Applied Physics, Stanford University, USA}

\begin{abstract}

A new method is presented for measuring terms in the elastoresistivity tensor $m_{ij}$ of single crystal samples with tetragonal symmetry. The technique is applied to a representative underdoped Fe-arsenide, Ba(Fe$_{0.975}$Co$_{0.025}$)$_2$As$_2$, revealing an anomalously large and anisotropic elastoresistance in comparison to simple metals. The $m_{66}$ coefficient follows a Curie-Weiss temperature dependence, providing direct evidence that the tetragonal-to-orthorhombic structural phase transition that occurs at $T_s$ = 97.5 K in this material is not the result of a true-proper ferro-elastic transition. Rather, the material suffers a pseudo-proper transition for which the lattice strain is not the primary order parameter.

\end{abstract}

\maketitle

\section{Introduction}

The elastoresistance of a material describes the relation between strain and changes in the electrical resistance. An important quantity in the semiconductor industry, for which strain effects must be carefully controlled \cite{Sun}, this property has been largely overlooked in the study of strongly correlated quantum materials. However, the elastoresistance tensor contains a wealth of information relating to both the symmetry of ordered states and also the nature of fluctuations. In addition, since electronically driven phase transitions often strongly affect the conductivity of a material, and since the order parameter must be coupled to the crystal lattice, such materials are likely to have anomalously large elastoresistance values relative to simple metals. In this paper we describe a new method to determine specific terms in the elastoresistivity tensor that is especially suitable for small samples, appropriate for typical cases of interest in the field of strongly correlated materials. We focus on the specific case of the underdoped iron arsenide Ba(Fe$_{0.975}$Co$_{0.025}$)$_2$As$_2$, for which an electronically driven structural phase transition leads to a divergence of specific terms in this tensor. 

The iron-arsenide superconductors undergo a tetragonal-to-orthorhombic structural transition at a temperature that either precedes or accompanies the onset of long range magnetic order. The origin of this effect has been discussed in various contexts, including orbital order \cite{CCLee_2009,Kruger_2009,Bascones_2010,Yin_2010,Lv_2010,CCChen_2010,Laad_2011}, a spin-driven nematic state \cite{Mazin_2009, Fang_2008, Xu_2008, Fernandes_2010, Fernandes_2012, Fernandes_Schmalian_2012, Daghofer_2010}, a combination of both \cite{Dagotto_2013}, and a Pomerancuk-type instability \cite{Lee_2009}. In a recent paper, we presented results of measurements of the induced resistivity anisotropy in the tetragonal state of the archetypal electron-doped iron-arsenide Ba(Fe$_{1-x}$Co$_x$)$_2$As$_2$ under conditions of uniaxial strain, using a piezoelectric stack to generate the strain \cite{JH_2012}. We related the induced in-plane resistivity anisotropy to the nematic susceptibility, and interpreted the divergence of this quantity as providing evidence that the structural transition is driven by an electronic nematic phase transition. In the current paper we describe the elastoresistive properties in greater detail, and show how terms in the elastoresistance tensor can be determined through measurement of longitudinal and transverse elastoresistance measurements. 

\begin{figure}
\includegraphics[width=8cm]{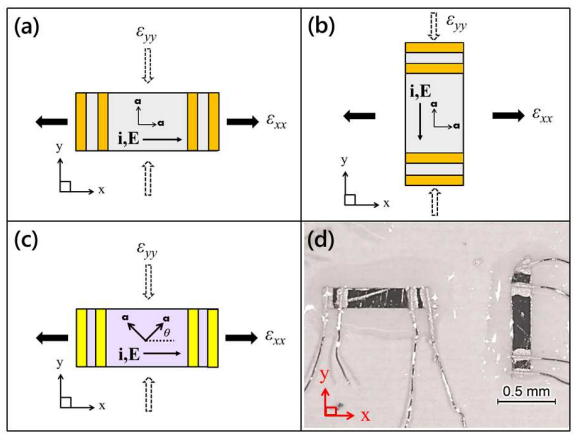} 
\caption{(Color online) Schematic diagrams illustrating measurement of (a) “longitudinal” elastoresistance (i.e. current $\parallel$ $\epsilon_{xx}$) and (b) “transverse” elastoresistance (i.e. current $\perp$ $\epsilon_{xx}$) for the specific case of $\epsilon_{xx}$ aligned along the [100]$_T$ tetragonal crystallographic direction. For a tensile strain in the x-direction, the strain in the y-direction is compressive and vice versa, with a ratio defined by the effective Poisson's ratio of the piezoelectric stack $\nu_p = -\epsilon_{yy}/\epsilon_{xx}$.  Panel (c) illustrates measurement of the longitudinal elastoresistance for $\epsilon_{xx}$ aligned along an arbitrary in-plane direction making an angle $\theta$ with respect to $[100]_T$. Strain, current directions and crystal axis orientation are indicated in all three panels. Gold stripes indicate current and voltage connections used for standard 4-terminal resistance measurements. Panel (d) shows a photograph of two representative crystals mounted on the surface of a PZT piezo stack for simultaneous measurement of the longitudinal (left crystal) and transverse (right crystal) elastoresistance. Scale bar indicates size of crystals, and red axes indicate crystal orientation.}
\label{Fig_1}
\end{figure}

The experiment itself is straightforward to describe. Single crystals of Ba(Fe$_{1-x}$Co$_x$)$_2$As$_2$ are glued to the top surface of a piezoelectric stack, and the strain varied while simultaneously measuring the induced changes in the resistance. Our initial experiments were inspired by the original work by Shayegan et al \cite{Shayegan_2003}, and followed a similar characterization of the piezoelectric stacks used for their measurements. Some care must be taken to ensure that the strain is fully transmitted to the sample, which can be readily checked by comparing strain measurements on the top surface of the sample relative to the surface of the piezoelectric stack. In our most recent experiments, described in this paper, samples are mounted in both longitudinal \textit{and} transverse geometries such that the strain $\epsilon_{xx}$ is parallel or perpendicular to the current in the sample respectively (illustrated in Figure 1). The strain $\epsilon_{xx}$ is modulated by varying the voltage applied to the piezoelectric stack (Figure 2(a)), and measured by strain gauges glued to the surface of the piezoelectric stack. The piezoelectric stack is characterized by an effective Poisson's ratio $\nu_p$ = $-\epsilon_{yy}/\epsilon_{xx}$, and as such the measurements are technically made under conditions of \textit{biaxial} in-plane strain, rather than uniaxial strain. However, the biaxial strain is highly anisotropic, having opposite signs for $\epsilon_{xx}$ and $\epsilon_{yy}$, so the part that couples to the bulk modulus ($\epsilon_{xx}$ + $\epsilon_{yy}$) is small compared to the part that couples to the orthorhombicity ($\epsilon_{xx}$ - $\epsilon_{yy}$). The effective Poisson's ratio of the piezoelectric stack can be readily measured using mutually transverse strain gauges, and is shown in Figure 2(b) as a function of temperature. For the purpose of this paper we follow a convention in which the value of $\nu_p$ is greater than one because of the choice of coordinate axes (an alternative convention would give values 1/$\nu_p$). 

\begin{figure}
\includegraphics[width=8.5cm]{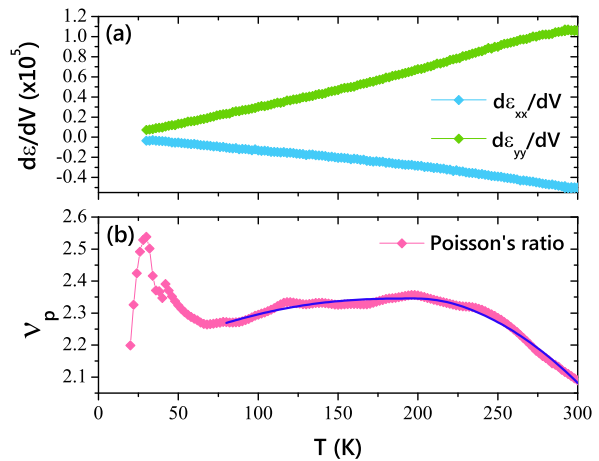} 
\caption{(Color online) (a) Temperature dependence of the strain per volt applied to the piezoelectric stack, $d\epsilon_{xx}/dV$ (blue) and $d\epsilon_{yy}/dV$ (green), measured using mutually orthogonal strain gauges. (b) Temperature dependence of the effective Poisson's ratio $\nu_p = -\epsilon_{yy}/\epsilon_{xx}$ of the piezoelectric stack, calculated using data shown in (a). Blue curve is an empirical fit of $\nu_p$ from 80K to 300K used for subsequent analysis of elastoresistance data.}
\label{Fig_1}
\end{figure}
    
\section{The elastoresistivity tensor}

\subsection{Definition of piezoresistivity and elastoresistivity tensors}
 
Elastoresistance is rarely discussed in the context of strongly correlated materials, although it conveys a wealth of important information relevant to understanding the nature of both fluctuations and also broken symmetry states. In the following paragraphs we briefly outline the tensor description of this quantity, making clear how specific terms in the elastoresistance tensor can be determined by a combination of longitudinal and transverse elastoresistance measurements made for specific crystal orientations. We start by describing the better known case of piezoresistance. 

The piezoresistance of a material relates changes in resistance ($R$) and the stresses experienced by the material. From the definition of resistivity ($\rho$), changes in the resistance are given by
\begin{equation}
\Delta R/R = \Delta\rho/\rho + \Delta L/L - \Delta A/A
\end{equation}
The first term on the right hand side of equation 1 describes changes in the resistivity of the material as a result of the applied stress. The second two terms describe purely geometric effects associated with changes in the length $L$ and cross-sectional area $A$. For typical metals, these geometric terms dominate the piezoresistance. However, as we show below, for the Fe-pnictides this is not the case, and we must also consider the effect of changes in the resistivity of the material as a function of applied stress.

The piezoresistivity is a fourth rank tensor, relating the applied stress and the resistivity, both second rank tensors. It is, however, more convenient to use the symmetry properties of the resistivity $\rho$ and stress $\tau$ tensors to express these as 6 component arrays, with components 
\begin{equation}
\tau = (\tau_{xx},\tau_{yy},\tau_{zz},\tau_{yz},\tau_{zx},\tau_{xy} )
\end{equation}
and
\begin{equation}
\rho = (\rho_{xx},\rho_{yy},\rho_{zz},\rho_{yz},\rho_{zx},\rho_{xy} )
\end{equation}
such that the piezoresistivity is described by a pseudo-second rank tensor $\pi$;
\begin{equation}
(\Delta\rho/\rho)_i = \displaystyle\sum_{k=1}^{6} \pi_{ik}\tau_k
\end{equation}
where 1 = $xx$, 2 = $yy$, 3 = $zz$, 4 = $yz$, 5 = $zx$, 6 = $xy$.

The stress can be expressed in terms of the elastic stiffness $C$ and the strain $\epsilon$,\cite{Strain}
\begin{equation}
\tau_k = \displaystyle\sum_{l=1}^{6} C_{kl}\epsilon_l
\end{equation} 
and hence we can readily derive an equivalent relation to equation 4, relating changes in the resistivity to the strains experienced by the material:
\begin{equation}
(\Delta\rho/\rho)_i = \displaystyle\sum_{k=1}^{6}(\displaystyle\sum_{j=1}^{6} \pi_{ij}C_{jk})\epsilon_k = \displaystyle\sum_{k=1}^{6} m_{ik}\epsilon_k
\end{equation}
Equation 6 defines the elastoresistivity tensor, sometimes also called the piezoresistive strain matrix, $m$. The measurements we describe below directly measure terms in this tensor. 
For a tetragonal material, appropriate for the Fe-arsenides for temperatures above the structural transition, there are 6 independent terms in the elastoresistivity tensor:
\begin{equation}
m_{ik} = \left(\begin{array}{cccccc}
m_{11} & m_{12} & m_{13} & 0 & 0 & 0 \\
m_{12} & m_{11} & m_{13} & 0 & 0 & 0 \\
m_{13} & m_{13} & m_{33} & 0 & 0 & 0 \\
0 & 0 &  & m_{44} & 0 & 0 \\
0 & 0 & 0 & 0 & m_{44} & 0 \\
0 & 0 & 0 & 0 & 0 & m_{66} \end{array}\right)
\end{equation}

\subsection{Measurement of terms in the elastoresistivity tensor}

For the longitudinal arrangement shown in Figure 1(a), if we neglect geometric factors, the change in resistance of the crystal is given by 
\begin{equation}
(\Delta R/R)_{xx} = m_{11}\epsilon_{xx} + m_{12}\epsilon_{yy} + m_{13}\epsilon_{zz}
\end{equation}
The strain in the $y$ direction, $\epsilon_{yy}$, is determined by the effective Poisson's ratio of the piezoelectric stack, $\nu_p$, whereas the strain in $z$ direction, perpendicular to the plane of the crystal, is given by an effective Poisson's ratio $\nu_s$. Although $\nu_s$ can be determined from knowledge of the elastic stiffness coefficients and $\nu_p$ \cite{Poisson}, we will soon see that it cancels out when we consider the difference of longitudinal and transverse values. Expressing the change of resistance in terms of just the strain in the $x$ direction, we then have for the longitudinal geometry (Figure 1(a)) 
\begin{equation}
(\Delta R/R)_{xx} = \epsilon_{xx}(m_{11} -\nu_p m_{12} -\nu_s m_{13})
\end{equation}
Similarly, the change in resistance for the transverse geometry (Figure 1(b)) is given by 
\begin{equation}
(\Delta R/R)_{yy} = \epsilon_{xx}(m_{12} -\nu_p m_{11} -\nu_s m_{13})
\end{equation}
To determine other relevant terms in the elastoresistance tensor, in particular $m_{66}$, we need to apply strain in directions that are not just along the crystallographic [100]$_T$ tetragonal axes. If we rotate the sample around the c-axis by an angle $\theta$, the elastoresistance tensor transforms \cite{Transformation} according to
\begin{equation}
m_{1'1'}= m_{11} - 2(m_{11} - m_{12} - 2m_{66}) \sin^2\theta \cos^2\theta
\end{equation}
\begin{equation}
m_{1'2'}= m_{12} + 2(m_{11} -m_{12} - 2m_{66}) \sin^2\theta \cos^2\theta  
\end{equation}
Hence, measurements of the longitudinal and transverse elastoresistance for samples rotated by an in-plane angle $\theta$ (illustrated in Figure 1(c)) are given by 
\begin{align}
\notag & (\Delta R/R)_{xx}  = \epsilon_{xx}(m_{1'1'} -\nu_p m_{1'2'} -\nu_s m_{1'3'}) \\ 
& = \epsilon_{xx}(m_{11} - 2(m_{11} - m_{12} - 2m_{66}) \sin^2\theta \cos^2\theta\\
\notag &-\nu_p (m_{12} + 2(m_{11} - m_{12} - 2m_{66}) \sin^2\theta \cos^2\theta) -\nu_s m_{1'3'})
\end{align}
Similarly, the transverse elastoresistance is given by
\begin{align}
\notag & (\Delta R/R)_{yy}  = \epsilon_{xx}(m_{1'2'} -\nu_p m_{1'1'} -\nu_s m_{1'3'}) \\ 
& = \epsilon_{xx}(m_{12} + 2(m_{11} - m_{12} - 2m_{66}) \sin^2\theta \cos^2\theta\\
\notag &-\nu_p (m_{11} - 2(m_{11} - m_{12} - 2m_{66}) \sin^2\theta \cos^2\theta) -\nu_s m_{1'3'})
\end{align}
It is useful at this stage to define the induced in-plane resistivity anisotropy, $N$, referred to a particular set of orthogonal in-plane axes x and y;
\begin{equation}
 N= \cfrac{\rho_{xx}-\rho_{yy}}{\cfrac{1}{2}(\rho_{xx}+\rho_{yy})} 
\end{equation} 
Since changes in the resistance have opposite signs for longitudinal and transverse configurations, to leading order the anisotropy is given by
\begin{equation}
 N \sim ((\Delta R/R)_{xx}-(\Delta R/R)_{yy}) 
\end{equation}
Hence, for arbitrary angle $\theta$, the induced anisotropy is given by
\begin{align}
\notag N = & \epsilon_{xx}(1+\nu_p)(m_{1'1'}-m_{1'2'})= \epsilon_{xx}(1+\nu_p)((m_{11}-m_{12})\\
& -4(m_{11} - m_{12} - 2m_{66}) \sin^2\theta \cos^2\theta)  
\end{align}
As anticipated, terms involving $\nu_sm_{13}$ cancel. For the two high symmetry cases $\theta$ = $0$ and $\pi/4$ (corresponding to $\epsilon_{xx}$ aligned along the tetragonal [100] and [110] directions respectively), we obtain
\begin{equation}
 N(\theta=0)=\epsilon_{xx}(1+\nu_p)(m_{11}-m_{12}) 
\end{equation}
\begin{equation}
 N(\theta=\pi/4)=\epsilon_{xx}(1+\nu_p)2m_{66}
\end{equation}
Hence, measurement of the induced resistance anisotropy, and in particular of the slope $dN/d\epsilon_{xx}$, combined with the measured effective Poisson's ratio of the piezoelectric stack, directly yields a measure of the coefficients $(m_{11}-m_{12})$ and $2m_{66}$ in the elastoresistivity tensor of the crystal sample \cite{2m66}.

\subsection{The elastoresistance of simple metals and semiconductors}

Microscopically, the resistivity of a metal is determined by a combination of Fermi surface parameters and the scattering rate. For the case of a single band, free electron model, the resistivity is given by the familiar expression $\rho = m^*/ne^2\tau$. In this case, the induced anisotropy due to the strain effects described above is given by 
\begin{equation}
 N= (\cfrac{\Delta\rho}{\rho})_{xx}-(\cfrac{\Delta\rho}{\rho})_{yy} = \cfrac{m^*_{xx}-m^*_{yy}}{m^*}-\cfrac{\tau_x-\tau_y}{\tau} 
\end{equation} 
where $m^*_{xx}$ and $m^*_{yy}$ (and $\tau_{x}$ and $\tau_{y}$) differ only because of the anisotropic biaxial strain. Hence, the measured elastoresistance and the derived slope $dN/d\epsilon_{xx}$ reflect induced anisotropy in both the effective mass $m^*$ and the relaxation time $\tau$
\begin{equation}
\cfrac{\partial N}{\partial\epsilon} = \cfrac{1}{m^*}\cfrac{\partial(m^*_{xx}-m^*_{yy})}{\partial\epsilon}-\cfrac{1}{\tau}\cfrac{\partial(\tau_x-\tau_y)}{\partial\epsilon}
\end{equation} 
For the case of a simple free-electron-like metal, anisotropy in $m^*$ and $\tau$ arise from strain-induced anisotropy in the bandwidth and phonon-spectrum respectively. Neither of these effects is large, and neither has a strong temperature dependence. For a multiband material, the situation is more complex, since the conductivity is determined from the sum of all of the pockets, each described by its own parameters. Even so, we can loosely think of the elastoresistivity as arising from some combination of strain-induced changes in the Fermi surface parameters, parameterized by an effective mass, and in the scattering, parameterized by some effective relaxation time. Measurement of the dc elastoresistivity coefficients cannot distinguish these effects, but equivalent measurements of the optical conductivity under applied strain are able to differentiate between anisotropy in scattering and spectral weight, as has recently been demonstrated for Ba(Fe$_{1-x}$Co$_x$)$_2$As$_2$ \cite{DeGiorgi_2011}.
 
If we also include the geometric factor described by equation 1, then the total change in the resistance as a function of strain yields the so called “gauge factor”, GF, of a material. (The name derives from the use of such materials as strain gauges: the gauge factor relates the change in resistance to the strain for a given configuration.) For the longitudinal configuration, this yields  
\begin{equation}
 GF = \cfrac{(\Delta R/R)_{xx}}{\epsilon_{xx}} = \cfrac{(\Delta \rho/\rho)_{xx}}{\epsilon_{xx}}+(1+2\nu)
\end{equation}
For ordinary metals, like copper, the first term on the right hand side is negligible as described above, and the gauge factor is almost solely determined by the geometric effect, yielding temperature-independent values with a magnitude close to 2 since the Poisson's ratio, $\nu$, for most metals is about 0.5. For the Fe-pnictides, the opposite is true, and the first term on the right hand side dominates for specific crystal orientations. This effect is intimately connected with the electronically-driven structural phase transition, as described in greater detail below. 

By way of comparison, for semiconductors, the gauge factor is typically dominated by the change of resistivity (the first term on the right side of equation 22)and consequently can be much larger than values obtained for typical simple metals. For example, the elastoresistivity coefficients for n-Si and p-Si, cubic semiconductors used in the most sensitive commercially available solid-state strain gauges, are respectively $m_{11}$ = -100.7, $m_{12}$ = 58.0, $m_{44}$ = -10.8 and $m_{11}$ = 9.5, $m_{12}$ = 1.7 and $m_{44}$ = 109.9 at room temperature \cite{Sun}, comparable to the maximum values observed for Ba(Fe$_{1-x}$Co$_{x}$)$_{2}$As$_{2}$ in this study. However, the physical mechanism that results in the large elastoresistance of semiconductors is very different, being associated with strain-induced changes in the band gap that strongly affect the majority carrier density. Consequently, the elastoresistance of semiconductors follows a characteristic $1/T$ temperature-dependence \cite{Sun, Kanda}. 

\section{Experimental Methods}

Single crystals of Ba(Fe$_{1-x}$Co$_{x}$)$_{2}$As$_{2}$ with $x$ = 0.025 were grown from a self-flux method as described elsewhere \cite{Mandrus_2008,JH_2009}. The composition was determined by electron microprobe analysis with an uncertainty of 0.0015. Crystals were cut into rectilinear bars with long sides having angles $\theta = 0^o$, $22.5^o$ and $45^o$ with respect to [100]$_T$ (i.e. the [100] direction referenced to the tetragonal crystal lattice), with an uncertainty less than $5^o$. Electrical contact was made to sputtered gold pads using Dupont 4929N silver paste, and the temperature-dependence of the resistance measured by a standard four-point technique. These samples were then glued to the top surface of a PZT piezoelectric stack (part No. : PSt150/5x5/7 cryo 1, from Piezomechanik GmbH) using five minute epoxy (from ITW Devcon) spread uniformly across the bottom and sides of each crystal. Care must be taken to minimize unintentional strain caused by the gluing, and can be best monitored by comparing the temperature-dependence of the resistance before ($R_0$)and after ($R(V=0)$) mounting on the piezoelectric. Two samples were mounted close together on the surface of the piezoelectric stack, as illustrated in Figure 1(d), enabling simultaneous measurement of the longitudinal and transverse elastoresistance for a given crystallographic orientation. The strain $\epsilon_{xx}$ was measured by a strain gauge glued on the other side of the piezoelectric stack. Elastoresistance measurements were made at fixed temperature by sweeping the voltage applied to the piezoelectric stack, typically between -50 V and + 150 V at a sweep rate of 8 V/s while simultaneously measuring the sample resistance and strain. Three complete hysteresis sweeps were made for each temperature, before the temperature was changed and allowed to stabilize ready for the next measurement.

\begin{figure}
\includegraphics[width=8.5cm]{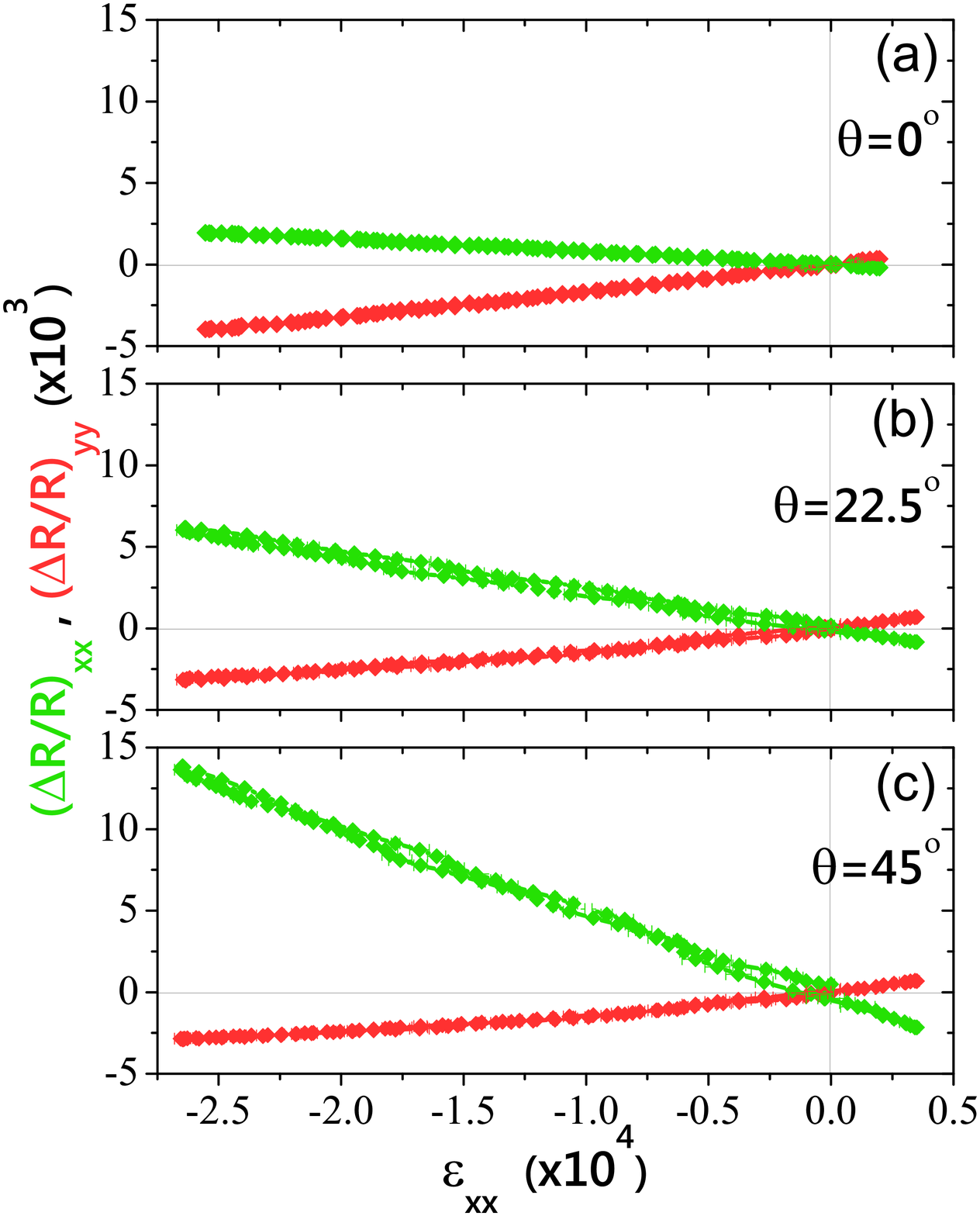} 
\caption{(Color online) Representative data showing the change in resistance as a function of strain $\epsilon_{xx}$ for longitudinal (red) and transverse (green) configurations for (a) $\theta = 0$ (i.e. $\epsilon_{xx}$ along $[100]_T$), (b) $\theta = 22.5^o$ and (c) $\theta = 45^o$ (i.e. $\epsilon_{xx}$ along $[110]_T$) . All measurements are at a temperature $T$ = 130 K.}
\label{Fig_3}
\end{figure}

\section{Results}

Representative longitudinal and transverse elastoresistance data ($(\Delta R/R)_{xx}$ and $(\Delta R/R)_{yy}$ respectively) as a function of strain $\epsilon_{xx}$ are shown in Figure 3 for a temperature of 130 K. In terms of the experimentally measured quantities, $\Delta R/R = (R(V)-R(V=0))/R_0$ and $\epsilon_{xx}= \epsilon_{xx}(V)-\epsilon_{xx}(V=0)$, where $R(V)$ and $\epsilon_{xx}$ are the resistance and strain measured at a given voltage $V$ applied to the piezo stack, and $R_0$ is the free-standing sample resistance. Equivalence to equation 13 and 14 follows from the linearity of $\Delta R/R$ with $\epsilon_{xx}$ (appropriate for small strains.) Data are shown for three distinct crystal orientations, corresponding to $\theta = 0^o$, $22.5^o$, $45^o$, where $\theta$ is the angle between the [100]$_T$ crystal axis and the $x$-axis (defined by the PZT piezo stack orientation- see Figure 1). Similar data were obtained at 2K increments between 80K and 280K. The measured elastoresistances vary linearly with the applied strain\cite{Nonlinear}, and are largest for $\theta = 45^o$. For each angle, the longitudinal and transverse elastoresistances are opposite in sign, but are not exactly equal in magnitude, as can be anticipated by inspection of equations 9 and 10. This effect arises from a combination of the effective Poisson’s ratio of the PZT piezo stack $\nu_p$ and also the effective Poisson’s ratio of the sample $\nu_s$, this latter quantity affecting strain in the $z$-direction. The sign of the elastoresistance for $\theta = 45^o$ is in accord with previous measurements of single crystals of Ba(Fe$_{1-x}$Co$_{x}$)$_{2}$As$_{2}$  held in a mechanical clamp, for which the longitudinal resistance increases under compressive stress (negative strain), while the transverse resistance decreases \cite{JH_2010, Tanatar_2010, Canfield_2010, HH_2011, Liang_2011, Blomberg_2011, Blomberg_2012, Ian_review}. Consequently the slope $d(\Delta R/R)_{xx}/d\epsilon_{xx}$ is negative while $d(\Delta R/R)_{yy}/d\epsilon_{xx}$ is positive. 

For each temperature, the induced anisotropy $N = (\Delta R/R)_{xx} - (\Delta R/R)_{yy}$ can be determined from the elastoresistance data; representative data based on the measurements shown in Figure 3 for $T$ = 130 K are shown in Figure 4 as a function of the strain $\epsilon_{xx}$ for each angle, $\theta = 0^o$, $22.5^o$ and $45^o$. The data are linear with strain, with intercept at zero anisotropy for zero applied strain. For all three angles the slope $dN/d\epsilon_{xx}$ is negative, as anticipated from Figure 3, indicating an increase (decrease) in the longitudinal  (transverse) elastoresistance as a function of strain $\epsilon_{xx}$. The largest slope is obtained for $\theta = 45^o$ (red data points in Figure 4). 

The anisotropy in $dN/d\epsilon_{xx}$ can be more clearly represented on a polar plot, shown in Figure 5, for which we have used the tetragonal crystal symmetry to generate equivalent data points. Data for $\theta = 0^o$ and $45^o$ were used to extract values for ($m_{11}-m_{12}$) and $2m_{66}$  respectively, following equations 18 and 19, and using measured values for $\nu_p$ shown in Figure 2(b). Following equation 17, these values can be used to calculate $dN/d\epsilon_{xx}$ for any angle $\theta$, shown by the solid blue line in Figure 5. As can be seen, the theoretical curve based on these values goes through the additional data point for $\theta = 22.5^o$. One could imagine making similar measurements for a more densely spaced range of angles, and using all of the data points to fit to equation 17 and hence obtain an even more precise estimates of ($m_{11}-m_{12}$) and $2m_{66}$, though for the small crystals used in this study that would be challenging. Even so, the data at the intermediate angle of $\theta = 22.5^o$ nicely confirm the anticipated angle-dependence based on the tensor transformation described in the previous section. 

The procedure used to determined ($m_{11}-m_{12}$) and $2m_{66}$ was repeated in 2K increments from 280 K down to 80 K (just below $T_s$). The T-dependence of these elastoresistivity coefficients is shown in Figure 6. Vertical lines in the figure indicate $T_N$ = 92.5 K and $T_s$ = 97.5 K, determined from resistivity measurements of the unstrained crystals before mounting on the piezoelectric stack (see for example ref [27]). The coefficient $2m_{66}$ increases as the temperature decreases, exhibiting a weak change in slope at $T_s$. In contrast ($m_{11}-m_{12}$) is smaller in magnitude, has a much weaker temperature dependence, and exhibits a sharper feature close to $T_s$ and $T_N$ which is presumably related to critical fluctuations.

\begin{figure}
\includegraphics[width=8.5cm]{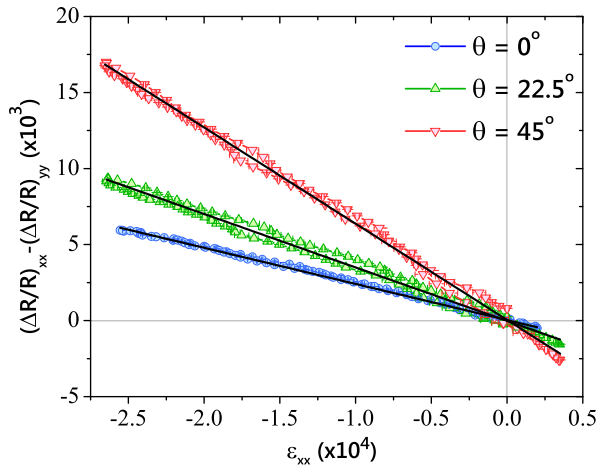} 
\caption{ (Color online) Representative data showing the induced anisotropy $N = (\Delta R/R)_{xx} - (\Delta R/R)_{yy}$ as a function of the strain $\epsilon_{xx}$ (see equations 16 and 17) at $T$ = 130K, for $\theta = 0^o$ (blue), $22.5^o$ (green) and  $45^o$ (red). Black lines show linear fits for each angle.}
\label{Fig_4}
\end{figure}
\begin{figure}
\includegraphics[width=7.5cm]{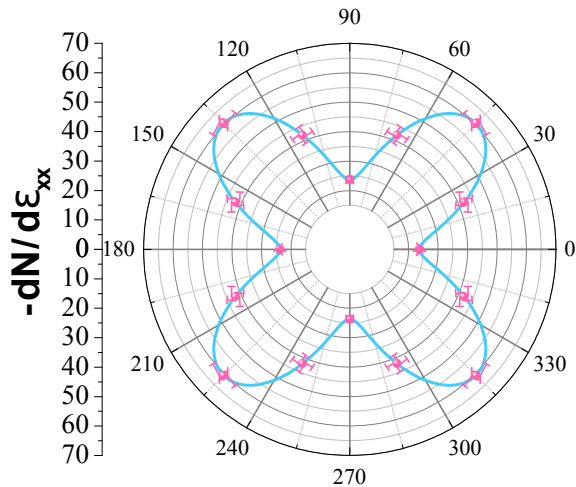} 
\caption{(Color online) Representative data showing the angle-dependence of the induced anisotropy $dN/d\epsilon_{xx}$ at $T$ = 130K. Solid points show data for $\theta = 0^o$, $22.5^o$ and  $45^o$, with additional equivalent data points generated by symmetry. Solid blue line shows calculated anisotropy based on equation 17 and measured ($m_{11}-m_{12}$) and $2m_{66}$ values.}
\label{Fig_5}
\end{figure}
\begin{figure}
\includegraphics[width=8.5cm]{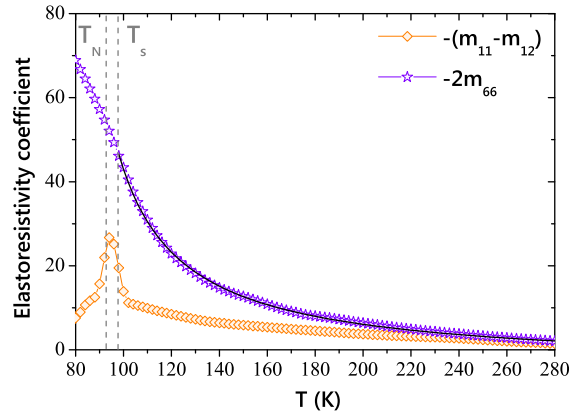} 
\caption{(Color online) Temperature-dependence of the elastoresistivity coefficients ($m_{11}-m_{12}$) and $2m_{66}$, determined from the induced anisotropy $dN/d\epsilon_{xx}$ for $\theta = 0^o$ and $45^o$ respectively. Black line shows fit to Curie-Weiss model for the $m_{66}$ coefficient; $2m_{66} = \lambda/[a_0(T-T^*)] + 2m_{66}^0$. Vertical dashed lines mark $T_s$ and $T_N$ of the sample.}
\label{Fig_6}
\end{figure}
\begin{figure}[h]
\includegraphics[width=8.5cm]{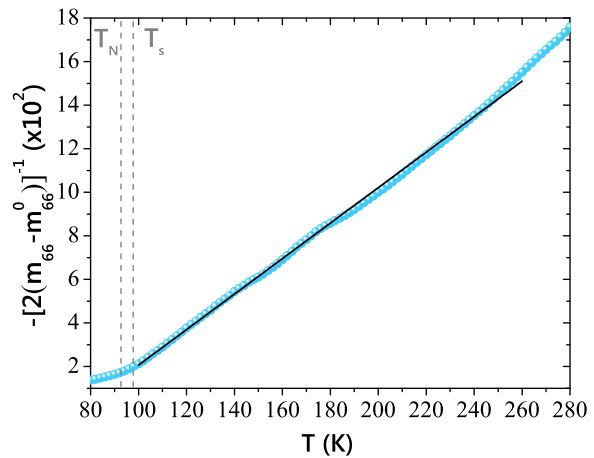} 
\caption{(Color online) Temperature-dependence of $[2(m_{66} - m_{66}^0)]^{-1}$, proportional to the inverse nematic susceptibility $\chi_N^{-1}$, for $\theta = 45^0$. Black line shows linear fit (Curie-Weiss model) to the data between 100 and 250 K. Vertical dashed lines mark $T_s$ and $T_N$ of the sample.}
\label{Fig_7}
\end{figure}

\section{Discussion}
\subsection{Comparison to simple metals}
Several features of the data shown in Figures 3-6 are unusual from the perspective of a simple metal. First, the magnitude of the elastoresistance is much larger than anticipated for a simple metal. For the specific cobalt concentration that we focus on in this paper, the $m_{66}$ coefficient reaches a maximum value of 24 ($2m_{66}=48$) for $T$ = 98K. The gauge factor for other doping levels rises to a maximum of 300 for $x$ = 0.0051 $\sim$ 0.007 \citep{JH_2012}. As described in section 2, for simple metals, one anticipates values of the gauge factor close to 2 since geometric effects dominate the elastoresistance. With the exception of the semimetals Sb and Bi, and ferromagnetic materials like Ni, this is indeed found to be the case \cite{Kuczynski}. 

Second, the temperature dependence of the elastoresistivity coefficient $m_{66}$ is found to diverge towards lower temperatures. Following our initial analysis in ref [18], the data can be very well fit by a Curie-Weiss temperature dependence;
\begin{equation} 
 2m_{66} = \cfrac{\lambda}{a_0(T-T^*)} + 2m_{66}^0
\end{equation}
The fit, shown by a solid line in Figure 6, yields fit parameters $\lambda/a_0 = -1238 \pm 46$K, $ T^* = 74 \pm 1$K, and $2m_{66}^0 = 3.6 \pm 0.3$. Using the $m_{66}^0$ value estimated from this fit, it is possible to plot the inverse elastoresistivity coefficient $-[2(m_{66} - m_{66}^0)]^{-1}$ as a function of temperature (Figure 7), yielding a clear linear behavior up to at least 250 K. While the strain-induced anisotropy for a simple metal will not necessarily be temperature-independent, there is no physical reason to anticipate a Curie-Weiss temperature dependence. 

Finally, the elastoresistivity exhibits a large anisotropy, not just in magnitude, but also in terms of the temperature dependence, neither of which effect is anticipated for a material described by a simple nearly-free-electron picture. 

In the following section, we describe how all of these observations can be understood in terms of electronic nematic order.

\subsection{Relation to the structural phase transition}

A natural order parameter for the ferroelastic structural phase transition is the spontaneous lattice strain that develops at $T_s$. The orthorhombic lattice parameters are rotated by $45^o$ with respect to the tetragonal unit cell, so referenced to the high-temperature tetragonal lattice the order parameter would be $\epsilon_6$ ($\epsilon_6 = \gamma_{ab}$ \cite{Strain}, where $a,b$ refer to the in-plane crystal axes). Softening of the associated elastic stiffness modulus $c_{66}$ in the tetragonal state has been observed via resonant ultrasound measurements \cite{Fernandes_2010, yoshizawa_structural_2012}. For a true-proper ferroelastic phase transition, the lattice strain is the primary order parameter, and the phase transition is driven by the elastic part of the free energy \cite{Ferroelasticity}. Hence, a Landau treatment of a true proper ferroelastic transition begins with a free energy expansion given by 
\begin{equation} 
 F= F_0 + \cfrac{c}{2} \epsilon^2 + \cfrac{d}{4} \epsilon^4 - h\epsilon
\end{equation} 
where $h$ represents an externally applied stress, the elastic modulus $c$ is assumed to have a temperature dependence $c = c_0(T-T_c)$, and we have dropped the tensor description for simplicity. Under conditions of zero stress, the material develops a spontaneous strain at $T = T_c$. Within such a picture, all other physical quantities, including the resistivity, develop an in-plane anisotropy at $T_c$ as a consequence of the orthorhombicity. Since all of these physical quantities share the same symmetry as the spontaneous strain, for small strains they are linearly proportional. Application of external stress in the same direction as the spontaneous strain (i.e. $\theta=45^o$) yields a finite strain for all temperatures, and the phase transition is smeared out. The associated strain-induced anisotropy in the resistivity in the tetragonal state is small, and hence linearly proportional to the strain, $\epsilon$. With reference to equation 19, the proportionality constant is given by the elastoresistivity coefficient $(1+\nu_p)2m_{66}$, the quantity that we have measured. Within the picture of a true-proper ferroelastic transition, $m_{66}$ is not necessarily temperature independent. For example, critical fluctuations very close to the structural transition could lead to anisotropic scattering, though this effect is anticipated to be small given that the orthorhombic distortion is a q=0 phenomenon (i.e. does not lead  to large momentum transfer). However, the observation of a divergent elastoresistivity coefficient $m_{66}$ that follows Curie-Weiss-like temperature dependence over a wide temperature range is completely inconsistent with a scenario in which the structural transition is driven by a true-proper ferroelastic transition. Rather, these data suggest either an improper or pseudo-proper ferroelastic transition, implying that the primary order parameter is not the elastic strain. 

For an improper ferroelastic transition, the primary order parameter has a different symmetry to the spontaneous strain. There is no reason based on any physical measurements to expect that this is the case for Ba(Fe$_{1-x}$Co$_{x}$)$_{2}$As$_{2}$. Rather, it is much more natural to assume that the primary order parameter shares the same symmetry as the spontaneous strain, which is indeed consistent with results of transport \cite{JH_2010, Tanatar_2010, Canfield_2010, HH_2011, Liang_2011, Blomberg_2011, Blomberg_2012, Ian_review}, and ARPES \cite{Ming_2011} measurements of detwinned crystals, and neutron scattering measurements \cite{Lester_2010, Li_2010, Harriger_2011} of twinned crystals. Lacking knowledge of the specific physical mechanism that is responsible for the electronic order that drives the lattice distortion, we need to choose suitably precise terminology that at least describes the broken symmetry, and therefore label this as an “electronic nematic” phase, following ref [44]. The only semantic caveats are; first, that in contrast to liquid crystal nematic phases, this electronic nematic phase breaks a discrete (rather than continuous) symmetry; and second, that use of the word nematic does not necessarily imply the orientational order of local objects.

Proceeding as described above, we define an electronic nematic order parameter, $\psi$. We do not directly measure this quantity, but assume that the resistivity anisotropy $N$ (equation 15) is a sensitive measure of it, and is linearly proportional for small enough values (appropriate in the tetragonal state for strained samples). As described in our previous paper \cite{JH_2012}, the associated free energy expansion is given by:
\begin{equation} 
 F= F_0 + \cfrac{a}{2} \psi^2 + \cfrac{b}{4} \psi^4 + \cfrac{c}{2} \epsilon^2 + \cfrac{d}{4} \epsilon^4 -\lambda \psi 
 \epsilon -h\epsilon
\end{equation}  
Since $\epsilon$ and $\psi$ have the same symmetry, they are coupled in a bilinear fashion. 

Furthermore, since strain acts as a field on the nematic order parameter via the coupling $\lambda$, the measured proportionality constant relating the resistivity anisotropy $N$ and the strain $\epsilon_6$ (i.e. the quantity $2m_{66}$ \cite{Nematic_sus}) is proportional to the “nematic susceptibility”:
\begin{equation}
\chi_N = \partial \psi/ \partial \epsilon \propto \partial N/ \partial \gamma_{ab} = 2m_{66}
\end{equation}
The temperature-dependence of the elastoresistivity coefficient $m_{66}$ can then be readily understood in terms of the pseudo-proper ferroelastic phase transition. In particular, minimization of the free energy (equation 25) with respect to both $\epsilon$ and $\psi$ for a strained sample yields 
\begin{equation} 
\chi_N = \cfrac{\lambda}{a}
\end{equation}  
Hence, the observation of an elastoresistivity coefficient $m_{66}$ that follows the mean-field Curie-Weiss form directly implies that the coefficient of the electronic nematic order parameter vanishes following $a = a_0(T - T^*)$, as anticipated if $\psi$ is the primary order parameter. The value $ T^* = 74 \pm 1$K, which is directly found from fitting the temperature dependence of the $m_{66}$ coefficient (Figure 6), would be the mean-field critical temperature for the nematic phase transition if there were no coupling to the crystal lattice. However, as we previously described in ref [18], the coupling $\lambda$ both ensures that there is a concurrent structural phase transition and also raises the critical temperature to a value 
\begin{equation} 
T_s = T^* + \cfrac{\lambda^2}{a_0c}
\end{equation} 
This is also consistent with the observation that $T^* < T_s$ ($T_s = 97.5 K$ for the specific cobalt concentration used in this study).

As a final comment we note that there is no evidence for any additional phase transition for temperature above $T_s$, up to our maximum measured temperature of $\sim$ 300K in contrast to recent claims \cite{Matsuda_2012}. However, the large magnitude of the elastoresistivity coefficients imply an extreme sensitivity to in-plane stress, and consequently care must be taken when interpreting results of experiments probing in-plane anisotropy for temperature above $T_s$ since residual strain can easily cause unintentional two-fold anisotropy.  

\section{Conclusion}

In writing this paper we have had two broad goals in mind. Our first goal has been to describe the elastoresistivity tensor, which relates changes in the resistance of a material to strain, and explain in some detail a new technique that we have developed to measure specific terms in this tensor. This is a physical quantity that has been largely neglected in the study of strongly correlated materials, but which can provide important insight to the nature of broken symmetry states and also fluctuations. The measurement technique is quite general, though our description and analysis has been specific to the case of a material with tetragonal symmetry. 

Our second goal has been to describe the angle- and temperature-dependence of the elastoresistance of the prototypical electron-doped iron arsenide Ba(Fe$_{0.975}$Co$_{0.025}$)$_2$As$_2$. Building on our earlier experiments \cite{JH_2012}, we have shown via a combination of longitudinal and transverse measurements that the $m_{66}$ elastoresistivity coefficient of this material follows a Curie-Weiss temperature dependence. This observation provides direct evidence that the elastic strain is not the primary order parameter, and hence that the tetragonal-to-orthorhombic structural phase transition should be classified as being a pseudo-proper ferroelastic phase transition driven by an electronic nematic instability. The divergent elastoresistivity coefficient can be related to the nematic susceptibility of the material, and hence bears witness to the presence of electronic nematic fluctuations in the normal state. 
 
\section{Acknowledgments}

The authors thank S. A. Kivelson, R. M. Fernandes, J. Schmalian, A. Shekhter and S. Raghu for helpful conversations. This work was supported by the DOE, Office of Basic Energy Sciences, under Contract No. DE-AC02-76SF00515.


\begin{thebibliography}{10}
\bibitem{Poisson} The effective Poisson's ratio $\nu_s$ is defined as $-\epsilon_{zz}/\epsilon_{xx}$, and can be expressed in terms of the elastic stiffness coefficient $C_{ij}$ : For a tetragonal system,        
\begin{equation}
\notag\left(\begin{array}{c}
\tau_1 \\
\tau_2 \\
\tau_3 \\
\tau_4 \\
\tau_5 \\
\tau_6 \end{array}\right) = \left(\begin{array}{cccccc}
C_{11} & C_{12} & C_{13} & 0 & 0 & 0 \\
C_{12} & C_{11} & C_{13} & 0 & 0 & 0 \\
C_{13} & C_{13} & C_{33} & 0 & 0 & 0 \\
0 & 0 &  & C_{44} & 0 & 0 \\
0 & 0 & 0 & 0 & C_{44} & 0 \\
0 & 0 & 0 & 0 & 0 & C_{66} \end{array}\right)\left(\begin{array}{c}
\epsilon_1 \\
\epsilon_2 \\
\epsilon_3 \\
\epsilon_4 \\
\epsilon_5 \\
\epsilon_6 \end{array}\right) 
\end{equation}
For our case, $\tau_3 = 0$ (there is no constraint is the z-direction), therefore 
\begin{equation}
\notag \begin{array}{l}
\mbox{$C_{13}\epsilon_1 + C_{13}\epsilon_2 + C_{33}\epsilon_3 = 0$} \\
\mbox{$C_{33}\epsilon_3 = -C_{13}(\epsilon_1 + \epsilon_2) = -C_{13}(1-\nu_p)\epsilon_1$} \\
\mbox{$ \nu_s= -\epsilon_3\ / \epsilon_1 = (1-\nu_p)C_{13}/C_{33} < 0$} \end{array}
\end{equation}
The result that $\nu_s < 0$ deserves a brief comment, since a free-standing crystal would have a positive Poisson's ratio. However, because the in-plane strain is determined by $\nu_p$, the effective Poisson's ratio of the piezoelectric stack, the strain in the z-direction is different to that of a free-standing crystal. In particular, since our choice of co-ordinate system yields $\nu_p = -\epsilon_2 / \epsilon_1 > 1$ (i.e. the strain in the y-direction is larger than that in the x-direction), the strain in the z-direction is opposite in sign to that of a free-standing crystal. Hence, a negative (compressive)strain in the x-direction yields a larger positive (tensile) strain in the y-direction (controlled by the effective Poisson's ratio of the piezoelectric stack, $\nu_p$), and therefore the strain in the z-direction (controlled by the effective Poisson's ratio,$\nu_s$) is negative (compressive).
\bibitem{Transformation} For a transformation of axes from one Cartesian system $(1,2,3)$ to another $(1',2',3')$, the piezoresistivity transforms as $\pmb{(\cfrac{\Delta \rho}{\rho})'}$= $\pmb{\alpha \cfrac{\Delta \rho}{\rho}}$, and the strain transforms as $\pmb{\epsilon'}$=$\pmb{R\alpha R^{-1} \epsilon}$ (here the strain is the engineering strain, following the definition in [20]), where $\pmb{\alpha}$ is the transformation matrix for a rotation of axes around the c-axis by an angle $\theta$, 
\begin{equation}
\notag\pmb{\alpha} = \left(\begin{array}{cccccc}
\cos^2\theta  & \sin^2\theta & 0 & 0 & 0 & 2\sin\theta \cos\theta  \\
\sin^2\theta & \cos^2\theta  & 0 & 0 & 0 & -2\sin\theta \cos\theta  \\
0 & 0 & 1 & 0 & 0 & 0 \\
0 & 0 & 0 & \cos\theta & -\sin\theta & 0 \\
0 & 0 & 0 & \sin\theta & \cos\theta & 0 \\
-\sin\theta \cos\theta & \sin\theta \cos\theta & 0 & 0 & 0 & \cos^2\theta-\sin^2\theta \end{array}\right)
\end{equation} 
and $\pmb{R}$ is a matrix used to take into account the factor of 2 between the engineering shear strain and the true shear strain (i.e. $\gamma_{xy}=2\epsilon_{xy}$). 
\begin{equation}
\notag\pmb{R} = \left(\begin{array}{cccccc}
1  & 0 & 0 & 0 & 0 & 0  \\
0 & 1  & 0 & 0 & 0 & 0  \\
0 & 0 & 1 & 0 & 0 & 0 \\
0 & 0 & 0 & 2 & 0 & 0 \\
0 & 0 & 0 & 0 & 2 & 0 \\
0 & 0 & 0 & 0 & 0 & 2 \end{array}\right),\qquad
\notag\pmb{R^{-1}} = \left(\begin{array}{cccccc}
1  & 0 & 0 & 0 & 0 & 0  \\
0 & 1  & 0 & 0 & 0 & 0  \\
0 & 0 & 1 & 0 & 0 & 0 \\
0 & 0 & 0 & \cfrac{1}{2} & 0 & 0 \\
0 & 0 & 0 & 0 & \cfrac{1}{2} & 0 \\
0 & 0 & 0 & 0 & 0 & \cfrac{1}{2} \end{array}\right)
\end{equation} 
Since $\pmb{(\cfrac{\Delta \rho}{\rho})'}$= $\pmb{m' \epsilon'}$ ($\pmb{m}$ is the elastoresistivity tensor), we finally obtain $\pmb{m'}$=$\pmb{\alpha m R \alpha^{-1} R^{-1}}$. The two coefficients $m_{1'1'}$ and $m_{1'2'}$ are expressed in the main text (equation 11 and 12). The coefficient $m_{6'6'}$ is given by $2\sin^2\theta \cos^2\theta(m_{11}-m_{12})+(\cos^2\theta-\sin^2\theta)^2 m_{66}$.
\bibitem{Nonlinear} The piezoelectric stack is made of layered PZT, and consequently 
the expansion is not perfectly homogeneous. Since the strain gauges that we use to characterize the behavior of the piezoelectric stack are relatively large, they are not sensitive to such inhomogeneous strains. However, the single crystal samples used in the study are considerably smaller (see figure 1(d)), and consequently can be more sensitive to the the local inhomogeneous expansion. Measurements of the elastoresistance for which the current is oriented along the PZT stacks are essentially unaffected by inhomogeneous strains, and are always perfectly linear. However, measurements of crystals with the current oriented in the transverse direction cross several PZT stacks and are therefore more sensitive to inhomogeneous strain, which sometimes leads to small deviations from perfect linearity. Since the data quality is better when a sample has current parallel to the PZT layers, not
only because of the more homogeneous expansion but also because 
the induced resistivity change is larger, it is natural to
define this configuration as longitudinal and the other configuration
as transverse. This choice of coordinate system ($x$-axis oriented along the PZT stack direction and $y$-axis oriented transverse to the stacks) yields the unconventional result that $\nu_p = -\epsilon_{yy}/\epsilon_{xx} > 1$.
\bibitem{Nematic_sus} The component of the nematic susceptibility that diverges for spontaneous nematic order in the [110]$_T$ direction is given by $\cfrac{\partial \psi}{\partial \gamma_{ab}}\propto \cfrac{\partial N}{\partial\gamma_{ab}}$. Referred to the experimental geometry ($x$ and $y$ defined by the piezoelectric stack and $\theta=45^o$ for measurement of $m_{66}$), this is equal to $\cfrac{1}{(1+\nu_p)}\cfrac{\partial N}{\partial \epsilon_{xx}}$, since $\gamma_{ab} = \epsilon_{xx}-\epsilon_{yy} = (1+\nu_p)\epsilon_{xx}$ in this case. The factor of $(1+\nu_p)$ arises because $\epsilon_{xx} \neq -\epsilon_{yy}$ and hence referred to the crystal axes the strain is not pure sheer. Hence, from equation 19 we obtain $\cfrac{\partial N}{\partial\gamma_{ab}}=2m_{66}$. Equivalently, for spontaneous nematic order in the [100]$_T$ direction, the component of the nematic susceptibility that diverges would be given by $\cfrac{\partial \psi}{\partial (\epsilon_{xx} - \epsilon_{yy})} \propto \cfrac{1}{(1+\nu_p)} \cfrac{\partial N}{\partial \epsilon_{xx}}=m_{11}-m_{12}$ for $\theta=0^o$ from equation 18. Hence, the two quantities $2m_{66}$ and $(m_{11}-m_{12})$ directly measure the components of the nematic susceptibility tensor relevant for spontaneous nematic order in the [110]$_T$ and [100]$_T$ directions respectively. In group theory, these correspond to the $B_{2g}$ and $B_{1g}$  irreducible representatives of the $D_{4h}$ point group with $xy$ and $x^2-y^2$ symmetry respectively.
\end{thebibliography}
\end{document}